# PROPAGATOR FOR THE FREE RELATIVISTIC PARTICLE ON ARCHIMEDEAN AND NONARCHIMEDEAN SPACES

*UDC: 530.1*

**D. D. Dimitrijević, G. S. Djordjević and Lj. Nešić**

Faculty of Science and Mathematics,
P.O. Box 224, 18001 Niš, Serbia and Montenegro

**Abstract.** *We consider dynamics of a free relativistic particle at very short distances, treating space-time as archimedean as well as nonarchimedean one. Usual action for the relativistic particle is nonlinear. Meanwhile, in the real case, that system may be treated like a system with quadratic (Hamiltonian) constraint. We perform similar procedure in $p$ -adic case, as the simplest example of a nonarchimedean space. The existence of the simplest vacuum state is considered and corresponding Green function is calculated. Similarities and differences between obtained results on both spaces are examined and possible physical implications are discussed.*

## 1. Introduction

It is generally believed that the notion of space-time as a continuous manifold should break down at very short distances of the order of the Planck length $\lambda_p \sim 10^{-35} m$ . This would arise from the process of measurement of space-time points based on quantum mechanics and gravity arguments [1]. Among many different possibilities for a mathematical background of a quantum theory on very small distances, noncommutative [2] and nonarchimedan [3] spaces appear as the most promising ones. The simplest example of a nonarchimedean space, and the most suitable mathematical machinery for describing it, comes from the theory of $p$-adic numbers and $p$-adic analysis.

Since 1987, there has been an important interest in application of $p$ -adic numbers in many branches of theoretical and mathematical physics (for a review, see [4]). Foundations of $p$ -adic quantum mechanics [4] and quantum cosmology [5] were worthwhile achivements in the attempt at deeper understanding of fundamental processes at very short distances. The adelic connection between ordinary and $p$ -adic quantum theory has been proposed for strings [6] and quantum mechanics [7]. Appearance of the nonarchimedean structure in frame of deformation quantization is also a very intriguing point [8].

It is very convenient to study some technically simpler cases as the free relativistic particle (FRP) on a nonarchimedean space. Generally, it is valuable for better understanding of physical processes at such ``exotic`` spaces. Also, it is known that relativistic quantum mechanics possesses many of the essential features of quantum cosmology. It is the one of the main reasons for further investigation of the FRP, including computation of the corresponding Green function. Although usual action for the FRP is nonlinear, as in the real case [9,10], that system may be treated like a system with quadratic constraint. This paper is concerned with $p$-adic aspects of such a system. Some attention is paid to the comparison with standard (over the field of real numbers) counterpart.

The paper is organised as follows. After the Introduction in Section 2 we recapitulate basic facts about $p$-adic mathematics. Section 3 is devoted to standard FRP. In Section 4 $p$-adic quantum mechanical propagator and conditions for the existence of the simplest vacuum state are briefly reviewed. The main result, $p$-adic counterparts of Green function are studied in Section 5 for all different $p$-adic cases. We complete this paper with remarks and conclusions concerning space-time structure from ordinary and $p$-adic, relativistic and non-relativistic point of view. Aspects of (non)adelicity in our approaches are briefly discussed.

## 2. $p$-ADIC MATHEMATICS

Let us note that all numerical experimental results belong to the field of rational numbers $Q$. The completion of this field with respect to the standard norm $|\cdot|_\infty$ (absolute value) leads to the field of real numbers $R \equiv Q_\infty$. According to the Ostrowski theorem, besides absolute value and $p$-adic norms ($p$ is a prime number) $|\cdot|_p$, there are no other non-equivalent and nontrivial norms on $Q$. The completion of $Q$ with respect to (a concrete prime number $p$) the $p$-adic norm leads to the (corresponding) $p$-adic number field $Q_p$.

Any $p$-adic number $x \in Q_p$, can be presented as an expansion [11]

$$x = x^\gamma (x_0 + x_1 p + x_2 p^2 + \cdots), \quad \gamma \in Z, \tag{1}$$

where $x_i$ are digits $x_i = 0,1,\cdots,p-1,\ x_0 \neq 0$. The $p$-adic norm of any term $x_i p^{\gamma+i}$ in (1) is $p^{-(\gamma+i)}$. The $p$-adic norm is nonarchimedean (ultrametric) one, i.e. $|x+y|_p \leq \max(|x|_p, |y|_p)$, and as a consequence, there are a lot of exotic features of $p$-adic spaces. This way, a choice of the acceptable formalism in quantum theory is constrained.

Generally speaking, there are two analyses over $Q_p$. One of them is connected with the map $\varphi: Q_p \to Q_p$ (mainly used in classical sector of the $p$-adic dynamics), and the second one is related to the map $\psi: Q_p \to C$ (in quantum dynamics).

In the case of mapping $\psi : Q_p \to C$, there is not standard derivative, and some types of pseudodifferential operators have been introduced [4,12,13]. However, it turns out that there is a well defined integral with the Haar measure. In particular, the Gauss integral will be employed

$$\int_{Q_p} \chi_p(\alpha x^2 + bx)dx = \lambda_p(\alpha)\,|\,2\beta\,|_p^{-1/2}\ \chi_p(-\tfrac{\beta^2}{4\alpha}),\ \alpha \neq 0\,, \tag{2}$$

where $\chi_p(u) = \exp(2\pi i\{u\}_p)$, an additive character on $Q_p$, is a complex-valued continuous function. Recall that in the real case one has $\chi_\infty(x) = \exp(-2\pi i x)$. Here, $\{u\}_p$ denotes the fractional part of $u \in Q_p$. By definition, $\lambda_p(\alpha)$ is an arithmetic complex-valued function [4]. Also, let us state the notation for the ring of $p$-adic integers, $p$-adic circle and disc, respectively:

$$\begin{aligned} &Z_p = \{x \in Q_p : |x|_p \le 1\}, \\ &S_\gamma(a) = \{x \in Q_p : |x-a|_p = p^\gamma\}, \quad \gamma \in Z\,, \\ &B_\gamma(a) = \{x \in Q_p : |x-a|_p \le p^\gamma\}, \quad \gamma \in Z, \quad \bigcup_{\nu=-\infty}^{\gamma} S_\gamma(a) = B_\gamma(a)\,. \end{aligned} \tag{3}$$

In the following, we will use the integral formula

$$\int_{S_\gamma} \chi_p(\varepsilon y)dy = \begin{cases} p^\gamma(1-p^{-1}), & if \quad |\varepsilon|_p \le p^{-\gamma} \\ -p^{\gamma-1}, & if \quad |\varepsilon|_p = p^{-\gamma+1} \\ 0, & if \quad |\varepsilon|_p \ge p^{-\gamma+2} \end{cases}\,. \tag{4}$$

Real and $p$-adic numbers can be considered simultaneously by concept of *adeles*. An adele [14] $a \in A$ is an infinite sequence $a = (a_\infty, a_2, \cdots, a_p, \cdots)$, where $a_\infty \in R$ and $a_p \in Q_p$ with the restriction that $a_p \in Z_p$ for all but a finite set $M$ of primes $p$. The set of all adeles $A$ can be written in the form

$$A = \bigcup_M A(M), \quad A(M) = R \times \prod_{p \in M} Q_p \times \prod_{p \notin M} Z_p\ . \tag{5}$$

$A$ is a topological space and it is a ring with respect to componentwise addition and multiplication.

## 3. Standard Free Relativistic Particle

The free relativistic particle belongs to a class of reparametrization-invariant theories, described by an action of the form [15]

$$S = \int_{t'}^{t''} dt[p_\alpha \dot{q}^\alpha - NH(p_\alpha, q^\alpha)]\,. \tag{6}$$

Here $N$ is Lagrange multiplier which enforces the constraint $H$=0.[1] The propagator for such (reparametrization-invariant) systems is represented by expression [15]

$$G(q^{\alpha''} \mid q^{\alpha'}) = \int dN(t''-t')\int Dp_\alpha Dq^\alpha \exp(i\int_{t'}^{t''} dt(p_\alpha \dot{q}^\alpha - NH)), \tag{7}$$

where we have an ordinary integration over $N$ and path integration over $p_\alpha$ and $q^\alpha$.

The functional integral part of (7) has the form of an ordinary quantum-mechanical propagator with the "time" $Nt$

$$\int Dp_\alpha Dq^\alpha \exp(i\int_{t'}^{t''} dt(p_\alpha \dot{q}^\alpha - NH)) = K(q''^\alpha, Nt'' \mid q'^\alpha, Nt'). \tag{8}$$

Introducing $T = N(t''-t')$, equation (7) becomes

$$G(q''^\alpha \mid q'^\alpha) = \int dTK(q''^\alpha, T \mid q'^\alpha, 0). \tag{9}$$

How can we see that FRP is a system of this type? The action for a FRP (in the flat configuration space) is usually written as

$$S = -mc\int_{\tau'}^{\tau''} d\tau \sqrt{\dot{x}^\mu \dot{x}^\nu \eta_{\mu\nu}}, \tag{10}$$

(where $\mu = 0,1,2,3$, and $\eta^{\mu\nu}$ is the usual Minkowski metric with signature $(-,+,+,+)$) and it is highly non-linear and unsuitable for quantum-mechanical investigations (a dot denotes a derivative with respect to $\tau$, where $\tau$ parametrizes worldline).

As we already said, a free relativistic particle can be treated as a system with the constraint [10,17]

$$\eta_{\mu\nu}k^\mu k^\nu + m^2c^2 = k^2 + m^2c^2 = 0, \tag{11}$$

which leads to the canonical Hamiltonian

$$H_c = N(k^2 + m^2c^2), \tag{12}$$

and to the Lagrangian

$$L = \dot{x}_\alpha k^\alpha - H_c = \frac{\dot{x}^2}{4N} - m^2c^2N, \tag{13}$$

where $\dot{x}_\alpha = \dfrac{\partial H_c}{\partial k^\alpha}$. Therefore, corresponding action for FRP is

$$S = \int_{\tau'}^{\tau''} d\tau \left[\frac{\dot{x}^2}{4N} - m^2c^2N\right]. \tag{14}$$

[1] The simplest examples of systems described by such action, beside FRP, are a nonrelativistic point particle in parametrized form, and minisuperspace models of quantum cosmology [16].

It is not too hard to prove that this action is invariant under reparametrization

$$\tau \to f(\tau), \quad i.e. \quad x(\tau) \to \tilde{x}(\tau) = x(f(\tau)) \,. \tag{15}$$

Namely, the infinitesimal version of (15) is

$$f(\tau) = \tau + \varepsilon(\tau) \,, \tag{16}$$

with corresponding change in the action

$$\delta S = -mc\sqrt{\dot{x}^{\mu}\dot{x}^{\nu}\eta_{\mu\nu}}\,\varepsilon\Big|_{\tau'}^{\tau''} \,, \tag{17}$$

which vanishes for arbitrary $\dot{x}(\tau'')$, $\dot{x}(\tau')$ if and only if

$$\varepsilon(\tau') = \varepsilon(\tau'') = 0 \,. \tag{18}$$

In the case of the proper time gauge ( $\dot{N} = 0$ ) the Green function (7) can be presented as

$$G(x'' \mid x') = \int dN(\tau'' - \tau') \int Dk\, Dx \exp(i\int_{\tau'}^{\tau''} d\tau[k\dot{x} - NH]) \,. \tag{19}$$

Redefining $T = N(\tau'' - \tau')$, this may be written explicitly [10]

$$G(x'' \mid x') = \frac{1}{(2\pi)^4}\int dT \int d^4k\ \exp(i[k(x'' - x') - T(k^2 + m^2c^2)]) \,. \tag{20}$$

If the range of $T$ is from $-\infty$ to $+\infty$, integration may be performed yielding

$$G(x'' \mid x') = \int \frac{d^4k}{(2\pi)^3}\ \delta(k^2 + m^2c^2) \exp(ik(x'' - x')) \,. \tag{21}$$

In this case, $G(x'' \mid x')$ is a solution of the Klein-Gordon equation

$$(-(\nabla''^2 - \frac{\partial^2}{c^2 \partial t''^2}) + m^2c^2) G(x'' \mid x') = 0 \,. \tag{22}$$

If the range of integration is $T \in [0, +\infty)$, then

$$G(x'' \mid x') = -i\int \frac{d^4k}{(2\pi)^4} \frac{\exp(ik(x'' - x'))}{(k^2 + m^2c^2 - i\varepsilon)} \,, \tag{23}$$

where factor $i\varepsilon$ ensures convergence at the upper end of the $T$ integration. In this case we get the Feynman Green's function, which satisfies

$$(-(\nabla''^2 - \frac{\partial^2}{c^2 \partial t''^2}) + m^2c^2) G(x'' \mid x') = -i\delta^{(4)}(x'' - x') \,. \tag{24}$$

## 4. Quantum Free $p$-adic Relativistic Particle

The above expresions from (12) to (14) are valid in the real case and they have place in the $p$-adic one. The model of quantum free $p$-adic relativistic particle was studied in details in Ref. 18. We will briefly recapitulate a few main lines from this article. The classical trajectory reads explicitly

$$\bar{x} = \frac{x''-x'}{\tau''-\tau'}\tau + \frac{x'\tau''-x''\tau'}{\tau''-\tau'}, \tag{25}$$

and the classical action is

$$\bar{S}(x'',T;x',0) = \frac{(x''-x')^2}{4T} - m^2c^2T\,. \tag{26}$$

Also, it should be noted that classical (real and $p$-adic) action maybe represented in the (4-dimensional) form

$$\begin{aligned}\bar{S} = &-\frac{(x''^0-x'^0)^2}{4T} - \frac{m^2c^2T}{4} + \frac{(x''^1-x'^1)^2}{4T} - \frac{m^2c^2T}{4}\\ &+\frac{(x''^2-x'^2)^2}{4T} - \frac{m^2c^2T}{4} + \frac{(x''^3-x'^3)^2}{4T} - \frac{m^2c^2T}{4} = \bar{S}^0 + \bar{S}^1 + \bar{S}^2 + \bar{S}^3\,,\end{aligned} \tag{27}$$

($x''$ and $x'$ are 4-vectors). Then, quantum mechanical propagator (kernel of the operator of evolution) might be factorised as a product of the four propagators

$$K_p(x'',T'';x',T') = \prod_{\mu=0}^{3} K_p^{\mu}(x''^{\mu},T'';x'^{\mu},T')\,. \tag{28}$$

Because action of the system is quadratic, the quantum mechanical propagator, for each index $\mu = 0,1,2,3$, has the form [19]

$$K_p^{\mu}(x''^{\mu},T'';x'^{\mu},T') = N_p^{\mu}(T'',T')\chi_p(-\frac{1}{h}\bar{S}^{\mu}(x''^{\mu},T'';x''^{\mu},T''))\,, \tag{29}$$

where $N_p^{\mu}(T'',T')$ is a normalization factor, $\bar{S}^{\mu}(x''^{\mu},T'';x'^{\mu},T')$ is a $p$-adic classical action (quadratic in respect to $x''^{\mu}$ and $x'^{\mu}$), and $h$ is the Planck constant. The $p$-adic quantum mechanical propagator (29) (as in the real case) has the property

$$\int_{Q_p} K_p^{\mu}(x''^{\mu},T'';x^{\mu},T)K_p^{\mu}(x^{\mu},T;x'^{\mu},T')Dx^{\mu} = K_p^{\mu}(x''^{\mu},T'';x'^{\mu},T')\,. \tag{30}$$

After integration and using general form of the normalization factor [18,19]

$$N_p(T'',T') = \left|N_p(T'',T')\right|_{\infty} A_p(T'',T')\,,$$

we have

$$K_p(x'',T''\mid x',T')=\frac{\lambda_p^2(4h(T''-T'))}{\left|2h(T''-T')\right|_p^2}\chi_p\left(-\frac{1}{h}\frac{(x''-x')^2}{4(T''-T')}+\frac{m^2c^2}{h}(T''-T')\right). \quad (31)$$

The vacuum state is such a functions $\psi_0$ for which

$$\int_{Q_p^4} K_p(x'',T''\mid x',T')\psi_0(x')d^4x_i=\psi_0(x'') \quad (32)$$

holds. Conditions for existence of the simplest $p$-adic vacuum state in the form $\Omega(|x|_p)=\begin{cases}1, & x\in Z_p\\ 0, & x\notin Z_p\end{cases}$, are:

1) for $p\neq 2$

$$\left|\frac{m^2c^2}{h^2}h(T''-T')\right|_p\le 1,\quad with\quad \left|\frac{m^2c^2}{h^2}\right|_p\le 1,\quad and\quad \left|h(T''-T')\right|_p\le 1, \quad (33)$$

2) for $p=2$

$$\left|h(T''-T')\right|_2<1,\quad and\quad \left|\frac{m^2c^2}{h^2}\right|_2\le 1. \quad (34)$$

We will return to this point in the conclusion.

## 5. $p$-ADIC PROPAGATOR IN CONFIGURATION SPACE (GREEN FUNCTION)

In article [18] the kernel of the operator of evolution for the FRP was calculated. Here we continue examination of this model by calculating corresponding Green function. In accordance with discussion in Section 3 and using the gauge $\dot{N}=0$, $p$-adic propagator in configuration space (Green function) is defined as

$$G_p(x''\mid x')=\int_{|hT|_p\le 1} dT K_p(x'',T\mid x',0), \quad (35)$$

i.e.,

$$G_p(x''\mid x')=\int_{|hT|_p\le 1} dT\frac{\lambda_p^2(4hT)}{|2hT|_p^2}\chi_p\left(-\frac{(x''-x')^2}{4hT}+\frac{m^2c^2}{h}T\right). \quad (36)$$

We note that for the $x'=0$, this expression leads to the appropriate $p$-adic Hartle-Hawking (HH) wave function. It underlines a necessity of calculation $p$-adic Green function for deeper understanding of the quantum model of FRP and possible application in quantum cosmology.

The nature of integration [4] enforces us to inspect three different cases:

$$\begin{aligned} &1)\quad p \equiv 1(\mathrm{mod})4, \quad \lambda_p^2(a) = 1, \\ &2)\quad p \equiv 3(\mathrm{mod})4, \quad \lambda_p^2(a) = \pm 1, \\ &3)\quad p = 2, \quad \lambda_p^2(a) = (-1)^{a_1} i. \end{aligned} \tag{37}$$

**5.1.** $p \equiv 1(\mathrm{mod})4$

As the range integration we choose $p$-adic ball $|hT|_p \le 1$. We introduce the substitutions

$$hT = z \Rightarrow dz = |h|_p dT, \quad z = \frac{1}{y} \Rightarrow dz = \frac{dy}{|y|_p^2} \Rightarrow dT = \frac{1}{|h|_p}\frac{dy}{|y|_p^2}, \quad q^2 = \frac{(x''-x')^2}{4}. \tag{38}$$

The Green function, using (4), is calculated

$$G_p(x''\,|\,x') = \frac{1}{|h|_p}\sum_{\gamma=0}^{+\infty}\int_{S_\gamma}\chi_p(-q^2 y)dy. \tag{39}$$

Let $\left|q^2\right|_p \ge p^2$, than $\left|q^2\right|_p \ge p^{-\gamma+2}, \quad \forall\gamma \ge 0, \quad \gamma \in N + \{0\}$. In respect to the third line of the (4) it follows that

$$\sum_{\gamma=0}^{+\infty}\int_{S_\gamma}\chi_p(-q^2 y)dy = 0 \Rightarrow G_p(x''\,|\,x') = 0, \quad for \quad \left|(x''-x')^2\right|_p \ge p^2. \tag{40}$$

Let now $\left|q^2\right|_p = p^\delta, \delta \in Z \setminus N$

$$\sum_{\gamma=0}^{+\infty}\int_{S_\gamma}\chi_p(-q^2 y)dy = \sum_{\gamma=0}^{-\delta} p^\gamma(1-p^{-1}) - p^{|\delta|} = -\frac{1}{p}. \tag{41}$$

That means

$$G_p(x''\,|\,x') = -\frac{1}{p}\frac{1}{|h|_p}, \quad for \quad \left|(x''-x')^2\right|_p \le p^\delta, \quad \delta \in Z \setminus N. \tag{42}$$

On the light cone *e.g.* $q^2 = 0$: $\sum_{\gamma=0}^{+\infty}\int_{S_\gamma} 1 dy = \sum_{\gamma=0}^{\infty} p^\gamma(1-p^{-1}) = \frac{1}{p}\sum_{\gamma=0}^{\infty} p^\gamma(p-1)$.

In the real case above expression would be regarded as **divergent** one. But keeping in mind that in $p$-adic case, for any $p$, holds $\sum_{\gamma=0}^{\infty} p^{\gamma}(p-1) = -1$, we represent $\sum_{\gamma=0}^{+\infty} \int_{S_\gamma} 1dy$ as

$$\sum_{\gamma=0}^{+\infty} \int_{S_\gamma} 1dy = -\frac{1}{p}.$$

This way, these results can be presented in a compact form

$$G_p(x'' \mid x') = \begin{cases} -\dfrac{1}{p|h|_p}, & if \quad \left|(x''-x')^2\right|_p \le p \\ 0, & if \quad \left|(x''-x')^2\right|_p > p \end{cases}. \tag{43}$$

**5.2.** $p \equiv 3(\mathrm{mod})4$

The only important difference in respect to $p \equiv 1(\mathrm{mod})4$ is

$$\lambda_p^2(4hT) = \lambda_p^2(z) = \lambda_p^2(y) = (-1)^{\gamma}, \quad |y|_p = p^{\gamma}. \tag{44}$$

Therefore, we calculate Green function as the sum of two parts (for detailed calculation see [20])

$$G_p(x'' \mid x') = |h|_p^{-1}\left(\sum_{\gamma=0}^{+\infty}(-1)^{2\gamma} \int_{S_{2\gamma}} dy \chi_p(-q^2 y)\right) + |h|_p^{-1}\left(\sum_{\gamma=0}^{+\infty}(-1)^{2\gamma+1} \int_{S_{2\gamma+1}} dy \chi_p(-q^2 y)\right). \tag{45}$$

Let $\left|q^2\right|_p = p^{\delta}$, for $\delta \ge 2$. It follows that

$$G_p(x'' \mid x') = 0, \quad for \quad |x''-x'|_p \ge p. \tag{46}$$

For $\delta = 1 \Rightarrow G_p(x'' \mid x') = \dfrac{-1}{p|h|_p}$, and for $\delta = 0 \Rightarrow G_p(x'' \mid x') = \dfrac{1}{|h|_p}(2-\dfrac{1}{p})$.

For odd $\delta$, $\delta < 0$

$$G_p(x'' \mid x') = |h|_p^{-1}\left(\sum_{\gamma=0}^{+\infty}(-1)^{\gamma} \int_{S_{2\gamma}} dy \chi_p(-q^2 y)\right) = |h|_p^{-1} \frac{1}{p+1}\left(\frac{p-1}{p} - 2p^{\delta+1}\right). \tag{47}$$

Similary, for even $\delta$, $\delta < 0$

$$G_p(x'' \mid x') = |h|_p^{-1}(p+1)^{-1}\left(\frac{p-1}{p} + 2p^{|\delta|+1}\right). \tag{48}$$

On the light cone, *i.e.* $q^2 = 0$, we find $G_p(x'' \mid x') = -\infty$. We can collect all the above results, for $p \equiv 3(\text{mod})4$ , in some more compact form.

$$G_p(x'' \mid x') = |h|_p^{-1} \begin{cases} 0, & if \quad \left|(x''-x')^2\right|_p \geq p^2 \\ -p^{-1}, & if \quad \left|(x''-x')^2\right|_p = p \\ 2-p^{-1}, & if \quad \left|(x''-x')^2\right|_p = 1 \\ (p+1)^{-1}\left(\frac{p-1}{p}+(-1)^{\delta}2p^{|\delta|+1}\right), & if \quad \left|(x''-x')^2\right|_p = p^{\delta}, \quad \delta<0 \\ -\infty, & if \quad \left|(x''-x')^2\right|_p = 0 \end{cases} .(49)$$

**5.3.** $p = 2$

In this case, Green function has the form

$$G_2(x'' \mid x') = 2|h|_2^{-1} \int_{|y|_2 \geq 2} dy \lambda_2^2(y) \chi_2(-q^2 y) . \tag{50}$$

For $y = 2^{-\gamma}(1+2y_1+\ldots)$, $\gamma \in N$, $\lambda_2^2(y) = (-1)^{y_1} i$ . Then

$$G_2(x'' \mid x') = |2h|_2^{-1} \sum_{\gamma=1}^{+\infty} \left(i \int_{S_\gamma,\, y_1=0} dy \chi_2(-q^2 y) - i \int_{S_\gamma,\, y_1=1} dy \chi_2(-q^2 y)\right) . \tag{51}$$

Let us compare [4] page 257-258

$$I_0 = \int_{S_\gamma,\, y_1=0} dy \chi_2(-q^2 y), \quad and \quad I_1 = \int_{S_\gamma,\, y_1=1} dy \chi_2(-q^2 y) . \tag{52}$$

1) If $q^2 = 0$, $I_0 = I_1 \Rightarrow G_2(x_j \mid x_i) = 0$ .
2) If $\left|q^2\right|_2 = 2^{\delta}$, $\delta \in N$, *and* $|y|_2 \leq 2^{-\delta}$, $I_0 = I_1$ .
3) If $\left|q^2\right|_2 = 2^{\delta}$, $\delta \in N$, *and* $|y|_2 = 2^{-\delta+1}$, $\{q^2 y\}_2 = \frac{1}{2}$, $I_0 = I_1$ .
4) If $\left|q^2\right|_2 = 2^{\delta}$, $\delta \in N$, *and* $|y|_2 = 2^{-\delta+2}$, $\{q^2 y\}_2 = \frac{1}{4}+\frac{1}{2}(y_1+q_1^2)$,
   $I_0 = (-i)^{q_1^2} 2^{-\delta} = -I_1$ .
5) If $\left|q^2\right|_2 = 2^{\delta}$, $\delta \in N$, *and* $|y|_2 \geq 2^{-\delta+3}$, $I_0 = I_1$ .

Omitting technical details [20] we present all those possibilities as

$$G_2(x''\,|\,x') = \begin{cases} 0, & if \quad q^2 = 0 \quad or \quad \left|(x''-x')^2\right|_2 \geq 1 \\ \dfrac{(-1)^{q_1^2}}{\left|h(x''-x')^2\right|_2}, & if \quad q^2 \neq 0 \quad and \quad \left|(x''-x')^2\right|_2 \leq \dfrac{1}{2} \end{cases} . \tag{53}$$

We can see that, practically in all cases, $G_p(x\,|\,0) \neq \Omega(x)$! It means that, except (43) i.e. case $p \equiv 1(\mathrm{mod})4$ , there is no vacuum state in the form of $\Omega(x)$ function (in Hartle-Hawking approach).

## 6. Conclusion

The vacuum state of $p$-adic quantum free relativistic particle (Section 4), like in non-relativistic case [4], is infinitelly degenerate. For example $\Omega(p^\gamma |x|_p)$, for $\gamma \in N + \{0\}$ are invariant functions in respect to the corresponding operator of evolution [18]. For the simplest vacuum state $\Omega(|x|_p)$ we find conditions (33) and (34). As we know, an adelic wave function must have following form

$$\psi(x) = \psi_\infty(x_\infty) \prod_{p \notin S} \Omega(p^\gamma |x|_p) \prod_{p \in S} \Omega(|x|_p) . \tag{54}$$

According to the definition of $\Omega$-function and $p$-adic properties of rational numbers we get

$$|\psi(x)|_\infty^2 = \begin{cases} |\psi_\infty(x)|_\infty^2, & x \in Z_p \\ 0, & x \in Q_p \setminus Z_p . \end{cases} \tag{55}$$

It would mean that probability to find free relativistic particle at some rational point is equal to that obtained in ordinary quantum theory if $x$ is an integer, but is equal to zero if $x$ is not an integer. Also we found $(T''-T') \in Z$, *i.e.* corresponding differences of proper time's points, for all $p$. $\gamma$'s that appears above represent multiples of the unity step of lattice.

It should be emphasized that in HH approach to the quantum cosmology the standard wave function of the universe is postulated to be [21]

$$\psi(x) = \int dTK(x,T\,|\,0,0) . \tag{56}$$

$p$-Adic and adelic HH wave function for a class of exactly solvable minisuperspace cosmological models was studied [22]

In this way, keeping in mind relations (9) and (35), we calculated propagator (Green function) $G_p(x''\,|\,x')$. The main results of the paper are contained in the formulas (43),

(49) and (53). We note that if we know $G_p(x\,|\,0)$ for a model, we also have information about the corresponding wave function, at least in frame of the HH approach.

It should be noted that in the quantum model of FRP discreteness (55) that can be obtained if we **give up** the minisuperspace/relativistic (HH) approach and neglect obtained $p$-adic Green functions. By physical implication, energy spectra related to the calculated Green functions and their interesting mathematical structure deserve further investigation. Rather significant differences between real and $p$-adic approach to the simple physical system, *e.g.* free relativistic particle might be suggested. Even in a different context, it is interesting to compare our results with those presented in Ref. [23]. For $p$-adic models in *field* theory see [24] and reference therein.

**Acknowledgement:** *The research of all three authors was partially supported by the Serbian Ministry of Science and Technology Project No 1643. We are thankful to B. Dragovich for useful discussion in the early phase of the preparation of this article. A part of this work was completed during a stay of G. Djordjević at LMU-Munich supported by DFG, and a stay at University of Freiburg supported by DAAD. G. Djordjević is very grateful to J. Wess and S. Waldmann for their kind hospitality. We are thankfull to the one of the referees for very valuable suggestions.*